# Machine Optics Studies for the LHC Measurements

Maciej Trzebiński[a]

[a]Institute of Nuclear Physics PAS, ul. Radzikowskiego 152, 31-342 Kraków, Poland

**ABSTRACT**

In this work the properties of scattered protons in the vicinity of the ATLAS Interaction Point (IP1) for various LHC optics settings are discussed. Firstly, the beam elements installed around IP1 are presented. Then the ATLAS forward detector systems: Absolute Luminosity For ATLAS (ALFA) and ATLAS Forward Protons (AFP) are described and their similarities and differences are discussed. Next, the various optics used at Large Hadron Collider (LHC) are described and the beam divergence and width at the Interaction Point as well as at the ATLAS forward detectors locations are calculated. Finally, the geometric acceptance of the ATLAS forward detectors is shown and the impact of the LHC collimators on it is discussed.

## 1. MOTIVATION

During the last few years an impressive number of results was obtained by all detectors installed at the Large Hadron Collider (LHC). They are ranging from the extensive tests of the Standard Model (with the certain highlight of the Higgs boson discovery[1,2]), to the constraints on hundreds of the so-called „beyond Standard Model" theories. This would not be possible without the LHC machine – technologically the most advanced and the biggest particle accelerator built so far[3]. This 27 km long synchrotron is designed to collide beams of protons with an energy up to 7 TeV and lead nuclei with an energy of 2.76 TeV per nucleon.

The outstanding performance of the LHC accelerator is a result of a careful beam preparation which follows the precise apparatus setting. It is worth stressing that the majority of measurements at the LHC are addressed to finding signs of new phenomena, which are expected to be very rare. As a consequence, in order to maximise the probability of observing such events, the collision rate needs to be maximised. On the other hand, there are processes like diffraction, for which a clean experimental environment is crucial. Moreover, the probability of having more than one proton-proton interaction during the beam crossing (the so called „pile-up", $\mu$) should be minimised, thus the standard LHC settings need to be modified.

The presence of an intact proton is a natural diffractive signature[4]. It is worth stressing that diffractive protons are usually scattered at very small angles (typically of the order of few microradians), so the detectors need to be placed far away (typically hundreds of meters) from the Interaction Point (IP) and as close to the beam as possible. It is also worth to know that since there are several magnets installed between the IP and the forward detectors, the proton trajectory is not a straight line. The setting of these magnets has a direct impact on the position of the proton at the forward detector location and, hence, the kinematic phase space available for the research.

In this paper the impact on the proton position at the forward detector location for a various optics settings in vicinity of ATLAS Interaction Point (IP1) is discussed. It is worth mentioning that the similar system of magnets and detectors exist at the CMS-TOTEM Interaction Point (IP5)[*].

---

E-mail: maciej.trzebinski@cern.ch

[*]With the minor differences as *e.g.* a crossing angle in $(x, z)$ rather than $(y, z)$ plane.

## 2. LHC OPTICS

The amount of data delivered by an accelerator is related to the luminosity, which can be calculated as:

$$L = \frac{n \cdot N_1 \cdot N_2 \cdot f}{4 \cdot \pi \cdot \sigma_x \cdot \sigma_y} F, \qquad (1)$$

where $N_1$ and $N_2$ are the number of particles per bunch in beam 1 and 2, correspondingly, $n$ is the number of bunches per beam, $f$ is the revolution frequency, $\sigma_x$ and $\sigma_y$ are the Gaussian sizes of the beams and $F$ is the geometric luminosity reduction factor due to the crossing angle at the Interaction Point:

$$F = \left(1 + \left(\frac{\theta_c^* \, \sigma_z^*}{2 \, \sigma^*}\right)^2\right)^{-1/2},$$

where $\theta_c^*$ is the crossing angle at the IP, $\sigma_z^*$ – the bunch length[†], and $\sigma^*$ – the transverse beam size at the IP. The crossing angle is introduced in order to avoid unwanted parasitic interactions, *i.e.* when the protons interact with each other away from the Interaction Point.

Taking into account the fact that the beam envelope is circular in the $(x, y)$ plane at the Interaction Point, one can put $\sigma = \sigma_x = \sigma_y$. Moreover, setting:

$$\sigma = \sqrt{\frac{\varepsilon \cdot \beta}{\gamma}},$$

where $\varepsilon$ is the beam emittance, $\beta$ is a betatron function and $\gamma$ is beam Lorentz factor, one can rewrite Eq.(1) as:

$$L = \frac{n \cdot N_1 \cdot N_2 \cdot f \cdot \gamma}{4 \cdot \pi \cdot \varepsilon \cdot \beta^*} F.$$

In terms of accelerator optics[5], the betatron function $\beta$ is a measure of the distance from a certain point to the one at which the beam is twice as wide. The lower is the value of the betatron function at the IP ($\beta^*$), the smaller is the beam size, thus the larger is the luminosity. During the standard data taking, the $\beta^*$ value was being changed from 11 m (the so-called *injection optics*, very early data taking in year 2010) to 0.6 m (the so-called *collision optics*, end of the proton-proton running periods in year 2012). The ultimate LHC goal is to lower this value to 0.4 m.

Apart from the betatron function, a very important parameter is the beam emittance, $\varepsilon$, which is a measure of the average spread in the position-momentum phase space. The LHC was designed to obtain $\varepsilon = 3.75\,\mu$m·rad, but due to the outstanding performance this value is about 2 $\mu$m·rad in an average run. In this paper the former value of the emittance is used in the calculations of the beam properties around the forward detectors, whereas the latter one is employed when the beam behaviour at the IP is computed. Such an approach is consistent with the one of the LHC machine group and the real experimental conditions.

### 2.1 Special LHC Optics Requirements

The luminosity value will be maximised during the most of the LHC operation time. In few runs addressed to the hard diffractive studies, the magnets settings are planned to be unchanged, but the proton population in a bunch will be decreased to keep the pile-up at reasonably low level[‡].

On the other hand, there are processes like the elastic scattering, which could not be measured in conditions mentioned above. Therefore, in order to study them, one needs to consider a special LHC machine settings, called the *high-$\beta^*$ optics*. These modifications include:

- high value of the betatron function, which implies a very low beam angular divergence at the IP (*cf.* Eq. (2)),

---

[†]In this paper asterisk (*) denotes values at the Interaction Point.
[‡]The optimal value of pile-up depends on the process, but one could assume a value of $\mu \sim 1$ in such runs.

- low intensity bunches, needed to minimise the intra-beam scattering effects and to avoid an additional smearing,

- small number of bunches, to operate without the crossing angle,

- parallel-to-point focusing – a special feature which causes that the protons scattered with the same angle are focused at the same point in the forward detector (in case of the discussed ALFA detectors such focusing occurs in the $(y, z)$ plane),

- phase advance of $\psi = \pi/2$, which is needed for simplifying the so-called *transport matrix* (see[6] for details).

It is worth mentioning that during the LHC runs in years 2010-2012 the TOTEM Collaboration observed that the beam background in the *high-$\beta^*$* optics is smaller than in the runs with the $\beta^* = 0.55$ m one[7]. This triggered a discussion to use *high-$\beta^*$* settings with higher number of bunches, larger intensities and non-zero crossing angle for the hard diffractive measurements.

**2.2 LHC Beamline in the Vincity of ATLAS Interaction Point**

The LHC collides two proton beams. They circulate in two horizontally displaced beam pipes which join into a common one about 140 m away from the Interaction Points. The beam performing the clockwise motion (viewed from above the ring) is called *beam1* and the other one *beam2*.

The LHC magnetic lattice in vicinity of the ATLAS IP is presented in Figure 1. The quadrupole magnets are labelled with the letter Q while the dipole ones with the letter D. The final focusing triplet (Q1 - Q2 - Q3) is positioned about 40 m away from IP1. Other quadrupoles (Q4, Q5 and Q6) are installed around 160 m, 190 m and 220 m from the ATLAS IP[8,9]. Between the IP1 and 240 m two dipole magnets, D1 at 70 m and D2 at 150 m, are installed. These magnets are used for beam separation.

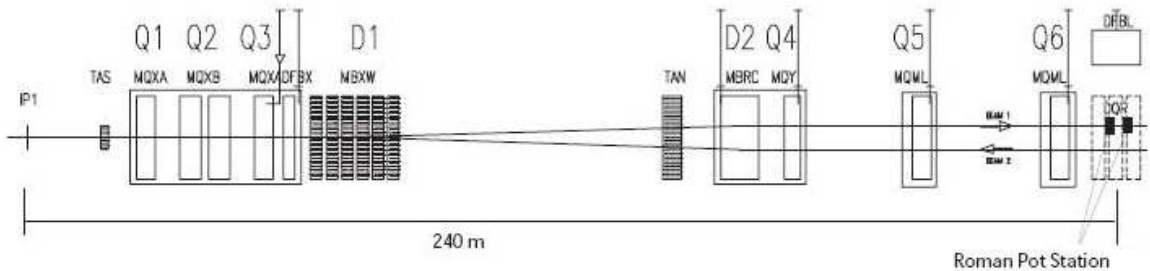

Figure 1. LHC magnet structure close to the ATLAS Interaction Point.

Apart from the main dipoles and quadrupoles there are several correction magnets used to compensate imperfections of the former ones. Moreover, there are several more LHC elements installed close to the ATLAS Interaction Point:

- Beam Position Monitors (BPM),

- Target Absorber Neutral (TAN) – absorber for neutral particles leaving the IP, located in front of the D1 dipole magnet on the side facing the ATLAS detector,

- Target Absorber Secondaries (TAS) – absorber for particles which could reach the quadrupole triplet. The first one is located in front of Q1, whereas the second one before the Q3 quadrupole magnet,

- TCL4 and TCL5 collimators which protect magnets from quenching. TCL4 is installed before D2 dipole whereas TCL5 before Q5 quadrupole magnet.

## 2.3 Forward Detectors

ATLAS is equipped with several forward detectors that monitor collision conditions, provide instantaneous luminosity estimates and measure particles scattered at small angles. There are three existing forward detectors[10]: LUminosity measurement using Cerenkov Integrating Detector (LUCID), Zero Degree Calorimeter (ZDC) and Absolute Luminosity For ATLAS (ALFA). There is also a plan to install the additional proton detectors – the ATLAS Forward Proton (AFP) stations[11]. The details of these detectors are described elsewhere (e.g.[6,10,11]), whereas here only the most important aspects of the forward proton tagges are discussed.

Diffractive protons are usually travelling close to the LHC beam. In order to measure them, there is a need to approach the beam as close as possible. One of the techniques which allow the detector insertion is the so-called „Roman Pot"[6]. The Roman Pot concept is based upon a detector volume (the pot) which is separated from the vacuum of the accelerator by a thin window and connected with bellows which allow the insertion into the beampipe. The design applied in the ATLAS forward detectors allows a precise positioning of the apparatus as close as 1 mm away from the beam centre.

The ALFA experimental set-up consists of four detector stations placed symmetrically with respect to the ATLAS IP at 237.4 m and 241.5 m. In each station there are two Roman Pot devices. The ALFA detectors were built using the scintillating fibre tracker technology, which allows the reconstruction of the proton position with a resolution of 30 $\mu$m. Since the detectors are planned to be used only during the special LHC runs, in which the instantaneous luminosity will be low, the applied technology is not a radiation-hard one.

Due to the design of the ALFA detectors it is impossible to use them during the standard LHC runs when the beam intensity is large. Therefore, the ATLAS Forward Proton (AFP) detectors are proposed to be installed 210 m far from the ATLAS Interaction Point.[11] Their designed proton position reconstruction resolution is 10(30) $\mu$m in the horizontal(vertical) direction. Moreover, in order to reduce the background originating from the pile-up, the AFP detectors are planned to be equipped with a fast timing system. The idea is to match the hard vertex reconstructed by the ATLAS central detector to the one reconstructed on a basis of the time of flight difference measured by the AFP stations. The foreseen timing reconstruction resolution is about 10 ps, which translates to the vertex reconstruction resolution of about 3 mm.

## 3. LHC BEAM PROPERTIES

The beam size at the ATLAS Interaction Point for various LHC optics and energies are listed in Table 1. These results were obtained using the MAD-X program[12,13] fed with the relevant LHC optics files[9].

Table 1. LHC beam transverse size at the ATLAS IP for various $\beta^*$ optics modes, proton energies and emittances.

| $\beta^*$ [m] | beam transverse size [mm] | | | | | |
| --- | --- | --- | --- | --- | --- | --- |
| | $E_{beam}$ = 3500 GeV | | $E_{beam}$ = 4000 GeV | | $E_{beam}$ = 7000 GeV | |
| | $\varepsilon$ = 2 $\mu$m·rad | $\varepsilon$ = 3.75 $\mu$m·rad | $\varepsilon$ = 2 $\mu$m·rad | $\varepsilon$ = 3.75 $\mu$m·rad | $\varepsilon$ = 2 $\mu$m·rad | $\varepsilon$ = 3.75 $\mu$m·rad |
| 0.55 | 0.017 | 0.024 | 0.016 | 0.022 | 0.012 | 0.017 |
| 90 | 0.22 | 0.30 | 0.21 | 0.28 | 0.16 | 0.21 |
| 1000 | 0.73 | 1.00 | 0.68 | 0.94 | 0.52 | 0.71 |
| 2625 | 1.20 | 1.60 | 1.10 | 1.50 | 0.84 | 1.15 |

For the measurement of the elastic scattering process, minimising the beam angular spread at the IP is a very important issue. This spread is calculated from the beam divergence:

$$\theta = 2\arctan\left(\frac{\sigma^* - \sigma_i}{2l}\right), \qquad (2)$$

where $\sigma^*$ is the width of the beam at IP, $\sigma_i$ – at a given point from the IP and $l$ is a distance between these two points. On this basis, the values of the beam divergence for various $\beta^*$ optics modes, proton energies and emittances were calculated. The results are summarised in Table 2.

The beam size at the forward detector location determines the minimum distance at which they are allowed to be inserted. Therefore, this knowledge is important for the event simulations and data analysis as it defines the kinematic regions that are accessible for a given optic settings. The results for the AFP and ALFA detector positions are listed in Tables 3 and 4. It is worth recalling that *beam1* and *beam2* are not identical, but the differences in their transverse size at the location of forward detectors are negligible.

Table 2. LHC beam divergence at the ATLAS IP for various $\beta^*$ optics modes, proton energies and emittances.

| $\beta^*$ [m] | beam divergence [MeV] | | | | | |
|---|---|---|---|---|---|---|
| | $E_{beam}$ = 3500 GeV | | $E_{beam}$ = 4000 GeV | | $E_{beam}$ = 7000 GeV | |
| | $\varepsilon$ = 2 $\mu$m·rad | $\varepsilon$ = 3.75 $\mu$m·rad | $\varepsilon$ = 2 $\mu$m·rad | $\varepsilon$ = 3.75 $\mu$m·rad | $\varepsilon$ = 2 $\mu$m·rad | $\varepsilon$ = 3.75 $\mu$m·rad |
| 0.55 | 107 | 146 | 114 | 156 | 151 | 206 |
| 90 | 1.00 | 1.38 | 1.07 | 1.47 | 1.42 | 1.95 |
| 1000 | 0.028 | 0.038 | 0.029 | 0.040 | 0.039 | 0.053 |
| 2625 | 0.0064 | 0.0088 | 0.0069 | 0.0094 | 0.0091 | 0.0125 |

Table 3. LHC beam transverse size in $x$ at the AFP and ALFA stations for various $\beta^*$ optics modes, proton energies and emittances.

| $\beta^*$ [m] | beam transverse size in $x$ [mm] | | | | | |
|---|---|---|---|---|---|---|
| | $E_{beam}$ = 3500 GeV | | $E_{beam}$ = 4000 GeV | | $E_{beam}$ = 7000 GeV | |
| | $\varepsilon$ = 2 $\mu$m·rad | $\varepsilon$ = 3.75 $\mu$m·rad | $\varepsilon$ = 2 $\mu$m·rad | $\varepsilon$ = 3.75 $\mu$m·rad | $\varepsilon$ = 2 $\mu$m·rad | $\varepsilon$ = 3.75 $\mu$m·rad |
| z = 204 m (AFP first station) | | | | | | |
| 0.55 | 0.20 | 0.27 | 0.19 | 0.26 | 0.14 | 0.19 |
| 90 | 0.61 | 0.83 | 0.57 | 0.78 | 0.43 | 0.59 |
| 1000 | 0.79 | 1.08 | 0.74 | 1.01 | 0.56 | 0.76 |
| 2625 | 0.35 | 0.48 | 0.33 | 0.45 | 0.25 | 0.34 |
| z = 212 m (AFP second station) | | | | | | |
| 0.55 | 0.14 | 0.20 | 0.13 | 0.18 | 0.10 | 0.14 |
| 90 | 0.51 | 0.69 | 0.47 | 0.65 | 0.36 | 0.49 |
| 1000 | 0.67 | 0.92 | 0.63 | 0.86 | 0.48 | 0.65 |
| 2625 | 0.31 | 0.42 | 0.29 | 0.40 | 0.22 | 0.30 |
| z = 237.398 m (ALFA first station) | | | | | | |
| 0.55 | 0.10 | 0.14 | 0.09 | 0.13 | 0.07 | 0.10 |
| 90 | 0.27 | 0.38 | 0.26 | 0.35 | 0.19 | 0.27 |
| 1000 | 0.39 | 0.54 | 0.37 | 0.50 | 0.28 | 0.38 |
| 2625 | 0.23 | 0.31 | 0.21 | 0.29 | 0.16 | 0.22 |
| z = 241.538 m (ALFA second station; till 2014) | | | | | | |
| 0.55 | 0.13 | 0.17 | 0.12 | 0.16 | 0.09 | 0.12 |
| 90 | 0.26 | 0.35 | 0.24 | 0.33 | 0.18 | 0.25 |
| 1000 | 0.37 | 0.51 | 0.35 | 0.48 | 0.26 | 0.36 |
| 2625 | 0.23 | 0.31 | 0.21 | 0.29 | 0.16 | 0.22 |
| z = 245.446 m (ALFA second station; after 2014) | | | | | | |
| 0.55 | 0.15 | 0.21 | 0.14 | 0.20 | 0.11 | 0.15 |
| 90 | 0.24 | 0.33 | 0.23 | 0.31 | 0.17 | 0.24 |
| 1000 | 0.35 | 0.48 | 0.33 | 0.45 | 0.25 | 0.34 |
| 2625 | 0.23 | 0.32 | 0.22 | 0.30 | 0.16 | 0.23 |

## 4. PROTON TRAJECTORIES

Proton trajectories are usually described in a curvilinear, right handed coordinate system (x, y, s). The local $s$-axis is tangent to the reference orbit at a given point of the beam trajectory. The two other axes are perpendicular to the reference orbit and are labelled $x$ (in the bent plane) and $y$ (perpendicular to the bent plane).

The trajectories in the nominal orbit trajectory reference frame for various LHC optics and different proton energies are presented in Figure 2. It can be observed that the proton deflection in the $x$-axis direction (outside the LHC ring) gets larger with decreasing proton energy *i.e.* with increasing proton relative energy loss:

$$\xi = 1 - E_{proton}/E_{beam}.$$

Table 4. LHC beam transverse size in $y$ at the AFP and ALFA stations for various $\beta^*$ optics modes, proton energies and emittances.

| $\beta^*$ [m] | beam transverse size in $y$ [mm] | | | | | |
|---|---|---|---|---|---|---|
| | $E_{beam}$ = 3500 GeV | | $E_{beam}$ = 4000 GeV | | $E_{beam}$ = 7000 GeV | |
| | $\varepsilon = 2$ $\mu$m·rad | $\varepsilon = 3.75$ $\mu$m·rad | $\varepsilon = 2$ $\mu$m·rad | $\varepsilon = 3.75$ $\mu$m·rad | $\varepsilon = 2$ $\mu$m·rad | $\varepsilon = 3.75$ $\mu$m·rad |
| z = 204 m (AFP first station) | | | | | | |
| 0.55 | 0.49 | 0.67 | 0.46 | 0.63 | 0.35 | 0.47 |
| 90 | 0.46 | 0.63 | 0.43 | 0.59 | 0.33 | 0.45 |
| 1000 | 0.16 | 0.22 | 0.15 | 0.21 | 0.12 | 0.16 |
| 2625 | 0.16 | 0.22 | 0.15 | 0.21 | 0.11 | 0.16 |
| z = 212 m (AFP second station) | | | | | | |
| 0.55 | 0.44 | 0.61 | 0.42 | 0.57 | 0.31 | 0.43 |
| 90 | 0.55 | 0.76 | 0.52 | 0.71 | 0.39 | 0.54 |
| 1000 | 0.19 | 0.26 | 0.18 | 0.24 | 0.13 | 0.18 |
| 2625 | 0.20 | 0.27 | 0.18 | 0.25 | 0.14 | 0.19 |
| z = 237.398 m (ALFA first station) | | | | | | |
| 0.55 | 0.29 | 0.40 | 0.27 | 0.38 | 0.21 | 0.28 |
| 90 | 0.68 | 0.93 | 0.63 | 0.87 | 0.48 | 0.66 |
| 1000 | 0.23 | 0.32 | 0.22 | 0.30 | 0.17 | 0.23 |
| 2625 | 0.26 | 0.35 | 0.24 | 0.33 | 0.18 | 0.25 |
| z = 241.538 m (ALFA second station; till 2014) | | | | | | |
| 0.55 | 0.27 | 0.37 | 0.25 | 0.34 | 0.19 | 0.26 |
| 90 | 0.65 | 0.89 | 0.60 | 0.83 | 0.46 | 0.63 |
| 1000 | 0.23 | 0.32 | 0.22 | 0.30 | 0.16 | 0.22 |
| 2625 | 0.25 | 0.34 | 0.23 | 0.32 | 0.18 | 0.24 |
| z = 245.446 m (ALFA second station; after 2014) | | | | | | |
| 0.55 | 0.24 | 0.33 | 0.23 | 0.31 | 0.17 | 0.23 |
| 90 | 0.62 | 0.85 | 0.58 | 0.79 | 0.44 | 0.60 |
| 1000 | 0.23 | 0.31 | 0.21 | 0.29 | 0.16 | 0.22 |
| 2625 | 0.24 | 0.33 | 0.23 | 0.31 | 0.17 | 0.24 |

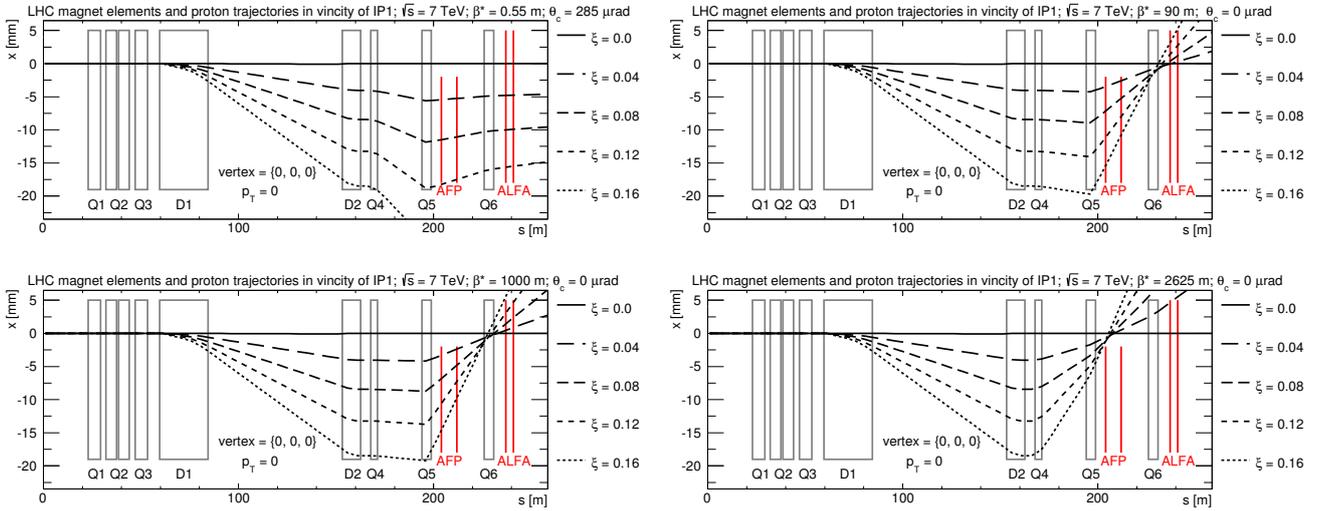

Figure 2. Energy dependence of the proton trajectory in $(x, s)$ plane for various LHC optics. Protons were generated at $(0, 0, 0)$ with transverse momentum $p_T = 0$. The crossing angle in horizontal plane was set to 142.5 $\mu$rad for the collision optics and equal to zero for the *high-$\beta^*$* ones. Trajectories were plotted for $\sqrt{s} = 7$ TeV, but are valid also for other energies as they scale with $\sqrt{s}$.

For the *high-β\** optics one can observe that for the distances above $s \approx 220$ m the orbits are bent inside the LHC ring, which is not the case for the collision optics. When the trajectory deflection is large enough (*cf.* trajectory with $\xi = 0.16$) the proton interacts with the LHC structures and is lost. One should also note that the trajectory reaches a maximum displacement from the nominal orbit at the distance between 160 m and 200 m. For the *high-β\** optics for $s > 220$ m the displacement increases with increasing value of the betatron function.

As can be seen in Figure 3, an introduction of the non-zero crossing angle causes that the proton trajectories are bent in the quadrupole triplet. In consequence, the proton position at the detector location in $(y, z)$ plane depends on its energy.

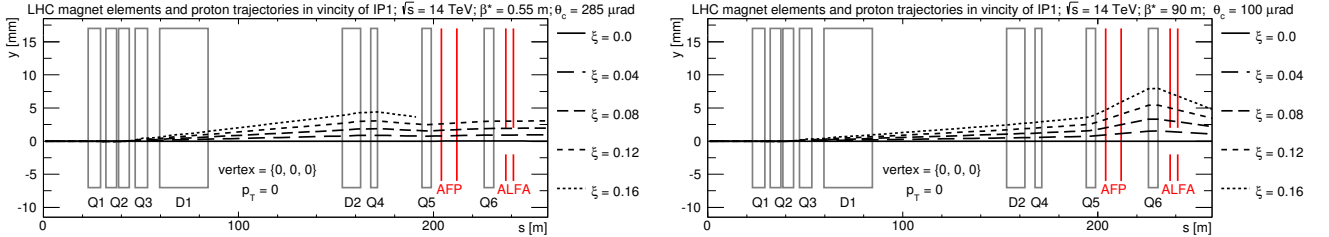

Figure 3. Energy dependence of the proton trajectory in $(y, s)$ plane for various LHC optics. Protons were generated at $(0, 0, 0)$ with transverse momentum $p_T = 0$. The crossing angle in horizontal plane was set to 142.5 $\mu$rad for the collision optics and to 50 $\mu$rad $\beta^* = 90$ m. Trajectories were plotted for $\sqrt{s} = 7$ TeV, but are valid also for other energies as they scale with $\sqrt{s}$.

The trajectories of protons scattered with the nominal energy and different values of the $p_x$ momentum component are plotted in Figure 4 for collision and *high-β\** optics. These trajectories depend on the beam energy, therefore all plots are done for $\sqrt{s} = 7$ and 14 TeV. As one can see the spread decreases with increasing $\sqrt{s}$.

In all presented optics the trajectories are focused behind the AFP detectors location. In the case of the *collision* and $\beta^* = 90$ m optics the maximum deviation from the nominal orbit is in the Q3 quadrupole. For $\beta^* = 1000$ m and $\beta^* = 2625$ m this maximum is around Q5 and Q4 magnet, respectively. It is worth noticing that this deviation increases with increasing value of the betatron function.

The proton behaviour at the nominal energy and different values of the $p_y$ momentum component for various LHC optics is shown in Figure 5. It should be stressed that the actual shape of the beam trajectory depends on its energy.

The maximum deviation in $y$ from the nominal orbit is in Q2 and Q6 for the collision and the *high-β\** optics, respectively. It is worth noticing that the presented LHC optics are designed in a way in which an increase of the $\beta^*$ value causes the increase of the deviation from the nominal orbit around ALFA detectors location. This in turn implies the possibility of measuring lower $p_y$ momenta which is of primordial importance for the ALFA physics programme[6].

## 5. GEOMETRIC ACCEPTANCE

For all measurements which are possible using the ATLAS forward detectors, it is important to understand the connection between the scattered proton energy and momentum and trajectory position in the detector. This dependence for various LHC optics is illustrated in Figures 6 and 7 for the AFP and ALFA detectors, respectively. In these figures the positions of elastically and diffractively scattered protons with various transverse momenta in the detector plane at the detector locations are shown. The detector shape is drawn using the solid lines.

From Figure 6 one can realise that protons scattered elastically ($\xi = 0$) will not reach the AFP detectors[§]. Protons scattered diffractively ($\xi \neq 0$) have negative value of $x$, *i.e.* they fly outside the LHC ring. The same behaviour is observed in case of the $\beta^* = 0.55$ m for ALFA stations (*cf.* Fig. 7). On the other hand, this picture changes for the *high-β\** optics as:

---

[§]Unless their $p_T$ is large enough which is possible but very improbable.

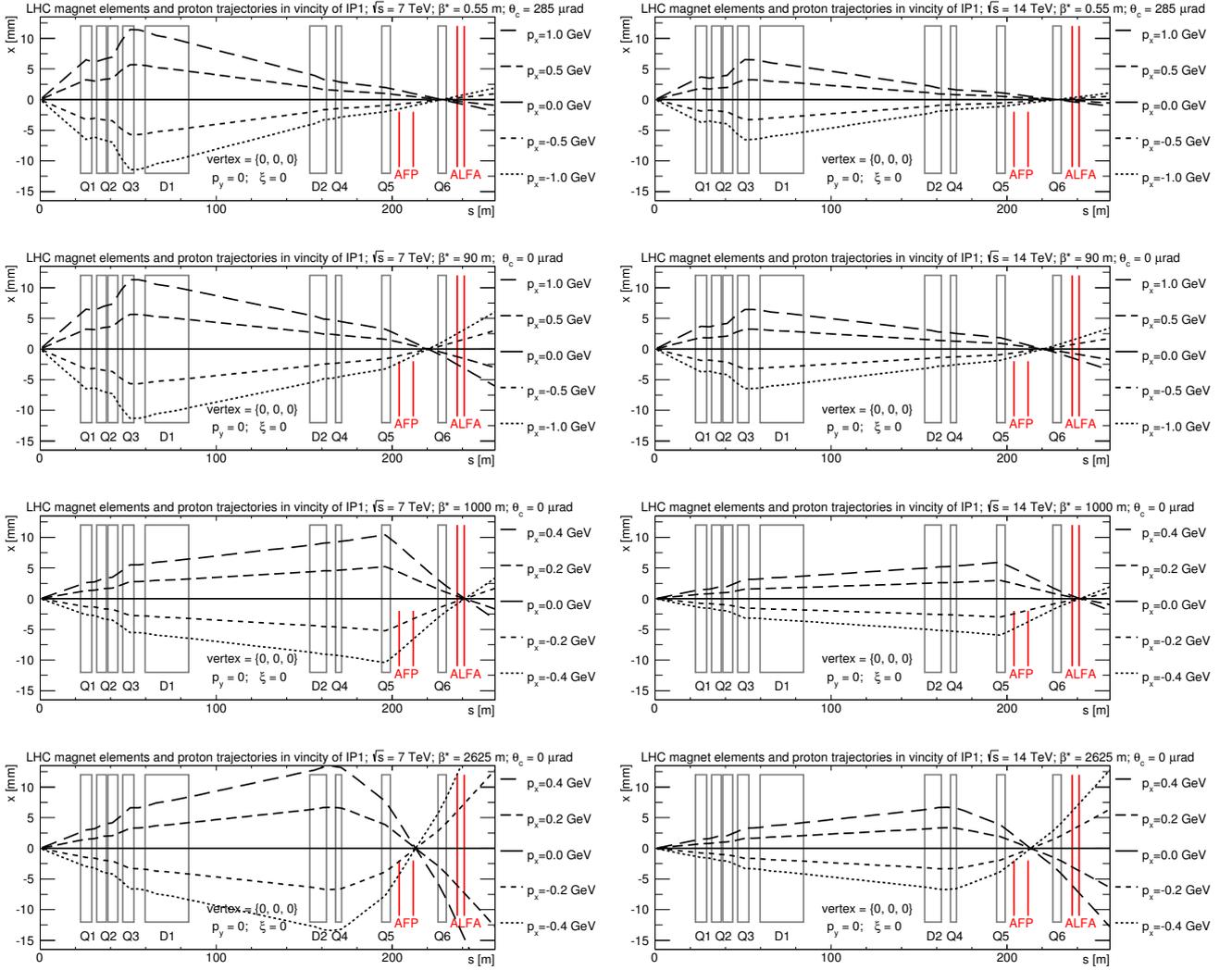

Figure 4. Transverse momentum dependence of the proton trajectory for various LHC optics for $\sqrt{s} = 7$ and 14 TeV. Protons were generated at $(0, 0, 0)$ with different $p_x$ momenta. The crossing angle in horizontal plane is set to 142.5 $\mu$rad for the collision optics and equal to zero for the *high-$\beta^*$* ones.

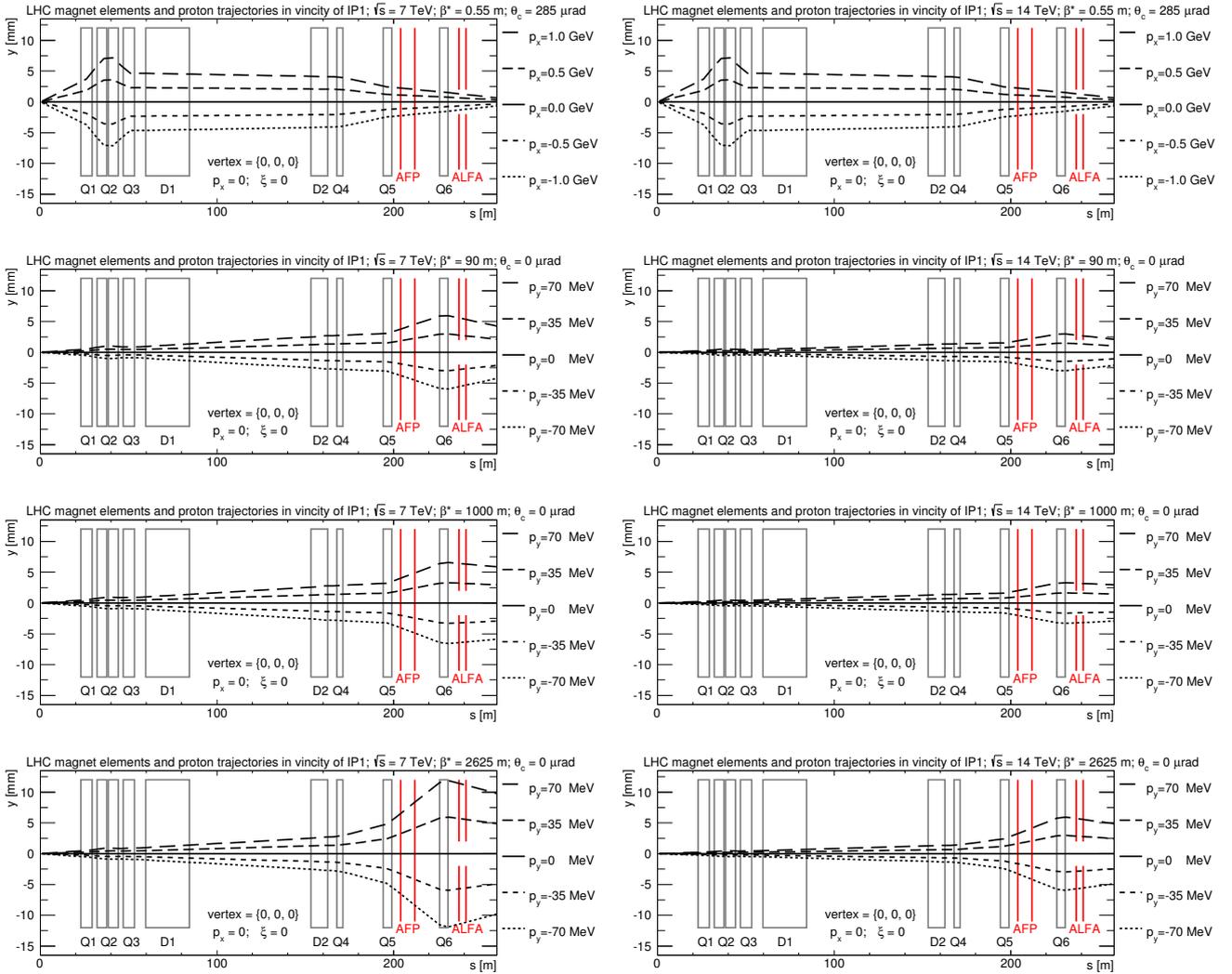

Figure 5. Transverse momentum dependence of the proton trajectory for various LHC optics for $\sqrt{s} = 7$ and 14 TeV. Protons were generated at $(0, 0, 0)$ with different $p_y$ momenta. The crossing angle in horizontal plane is set to 142.5 $\mu$rad for the collision optics and equal to zero for the *high-$\beta^*$* ones.

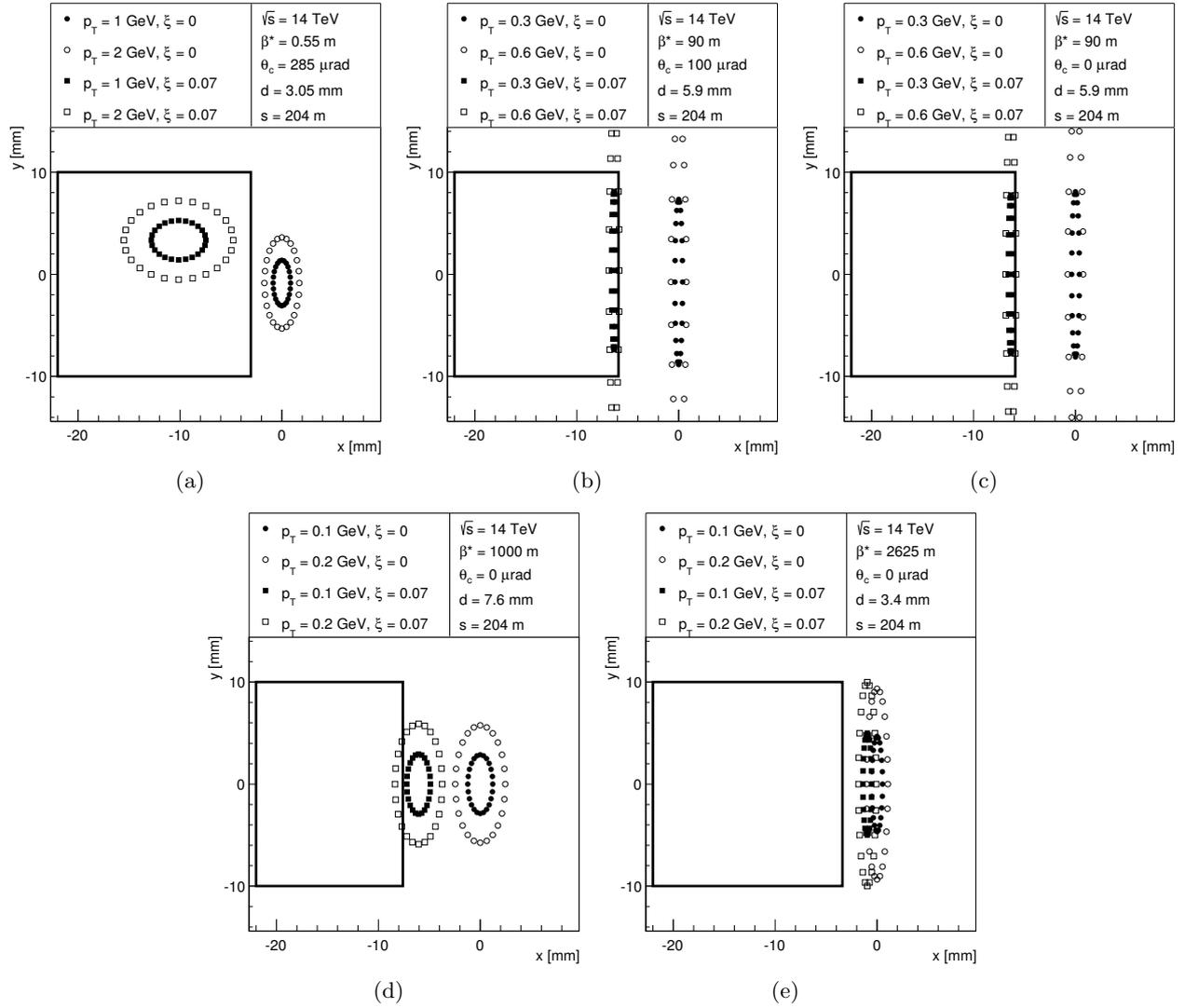

Figure 6. Proton positions with different relative energy loss ($\xi$) and transverse momentum ($p_T$) at the first AFP station for the different LHC optics. The solid lines mark the AFP detector active area. The distance from the beam centre is set to 15 $\sigma$ for the collision optics and 10 $\sigma$ for *high-$\beta^*$* ones.

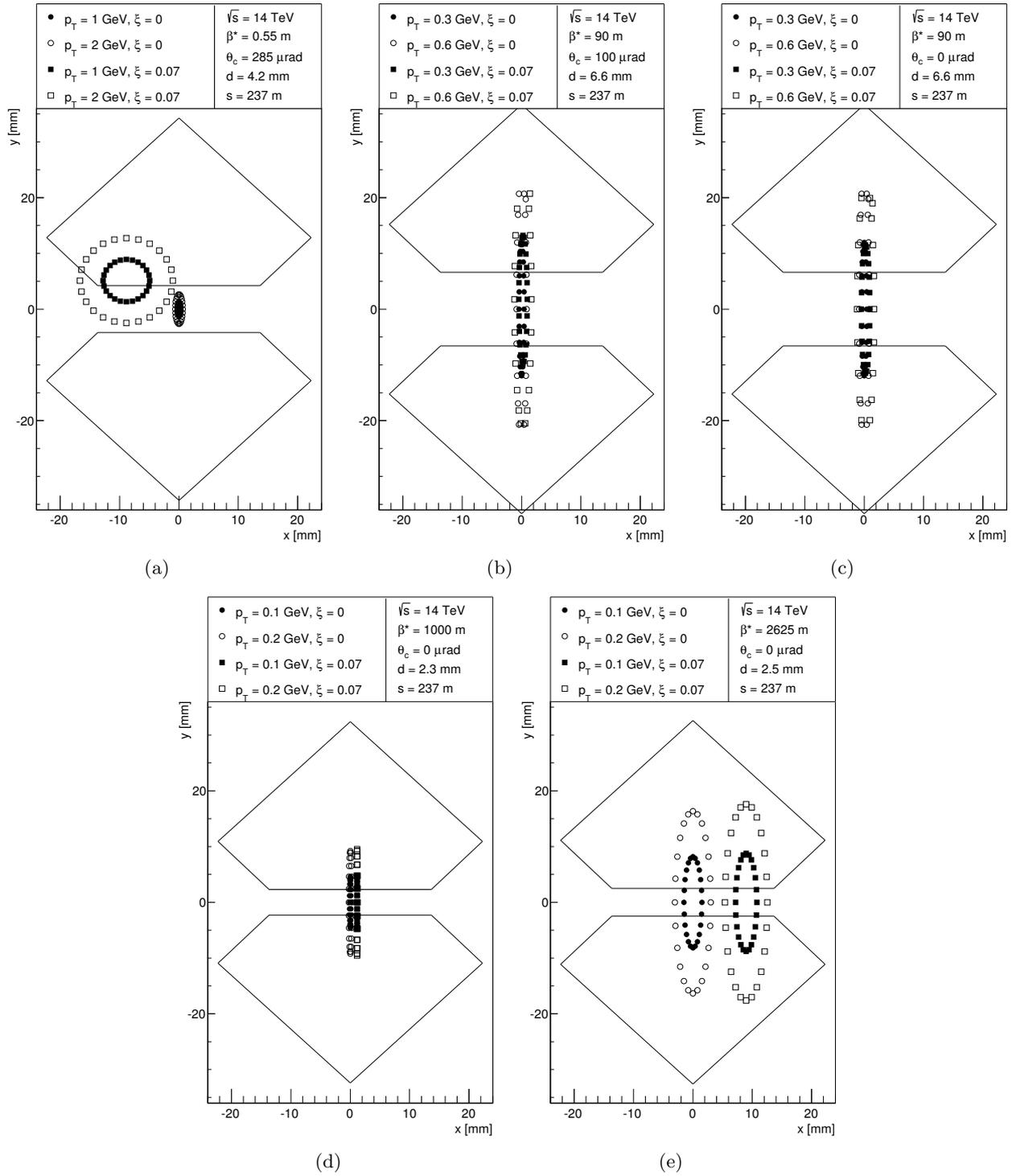

Figure 7. Proton positions with different relative energy loss ($\xi$) and transverse momentum ($p_T$) at the first ALFA station for the different LHC optics. The solid lines mark the ALFA detector active area. The distance from the beam centre is set to 15 $\sigma$ for the collision optics and 10 $\sigma$ for *high-$\beta^*$* ones.

- the access to lower values of transverse momentum increases with increasing $\beta^*$,
- trajectories of diffractive protons are bent towards the LHC ring centre and observed at positive $x$ values,
- the impact of the $p_y$-momentum component at the IP on the proton position at the detector plane is much larger than that due to $p_x$.

The geometric acceptance is defined as the ratio of the number of protons with a given relative energy loss ($\xi$) and transverse momentum ($p_T$) that reached the detector to the total number of the scattered protons having $\xi$ and $p_T$. Obviously, not all scattered protons can be measured in forward detectors as they can be too close to the beam to be detected or can hit the LHC element (a collimator, the beam pipe, a magnet) upstream the detector. In the presented calculations the following factors were taken into account:

- the beam properties at the IP,
- the beam chamber geometry,
- the LHC lattice magnetic properties,
- the distance between the beam centre and the detector edge.

The geometric acceptance of the first AFP station (located at 204 m from the IP1) is shown in Figure 8. The distance from the beam centre was set to 15 $\sigma$ for the collision optics and to 10 $\sigma$ for the *high-$\beta^*$* ones (*cf.* Table 3) and a 0.3 mm of dead material was added. It is worth noticing that since the AFP detectors approach the beam in the horizontal plane, only the $x$ width is meaningful.

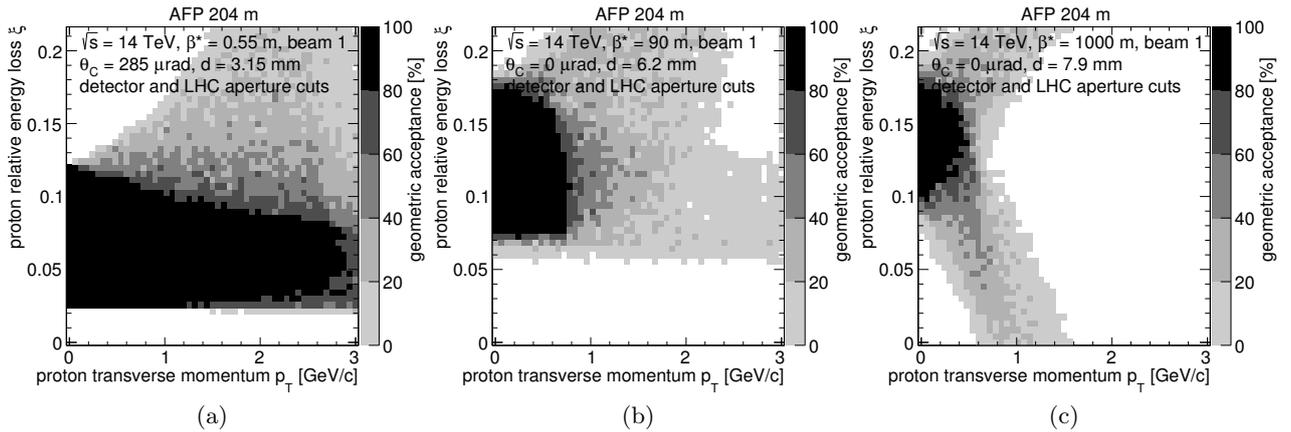

Figure 8. Geometrical acceptance of the AFP detectors as a function of the proton relative energy loss ($\xi$) and its transverse momentum ($p_T$) for various LHC optics settings. The beam properties at the IP, the beam chamber and the detector geometries, the distance between the detector edge and the beam centre were taken into account. This distance was set to 15 $\sigma$ for collision optics and to 10 $\sigma$ for *high-$\beta^*$* ones ($\varepsilon = 3.75$ $\mu$m·rad was used) and a 0.3 mm of dead material was assumed.

For the *collision* optics the region of high acceptance ($> 80\%$) is limited by $p_T < 3$ GeV and $0.02 < \xi < 0.12$. These limits change to $p_T < 1$ GeV and $0.07 < \xi < 0.17$ and $0.1 < \xi < 0.17$ for $\beta^* = 90$ and 1000 m optics, correspondingly. It is worth mentioning that the acceptances for $\beta^* = 90$ m with and without crossing angle are very similar. On the other hand, there is no acceptance in case of $\beta^* = 2625$ m (which is not surprising since for this optics the beam is strongly focused around 210 m – *cf.* Fig. 2d).

The results for the first ALFA station (located 237 m from the IP1) are shown in Figure 9. Also in this case the distance from the beam centre was set to 15 $\sigma$ for the collision optics and to 10 $\sigma$ for the *high-$\beta^*$* ones (see Table 4) and a 0.3 mm of dead material was added.

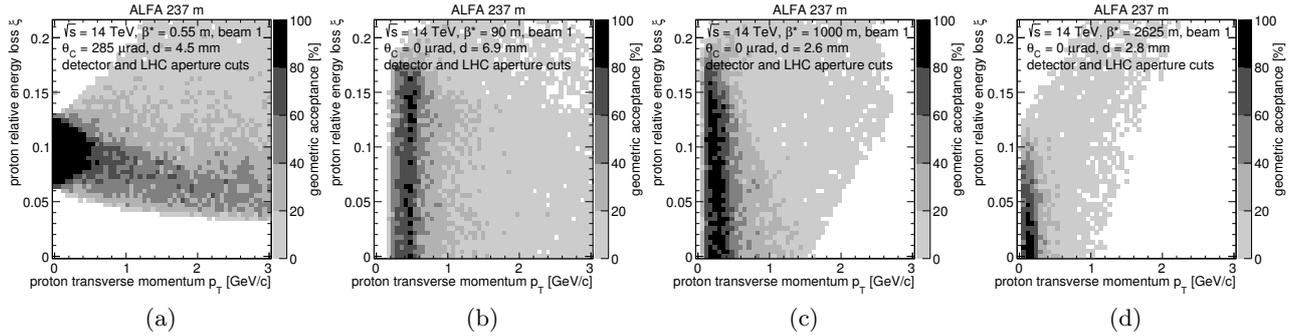

(a) (b) (c) (d)

Figure 9. Geometrical acceptance of the ALFA detectors as a function of the proton relative energy loss ($\xi$) and its transverse momentum ($p_T$) for various LHC optics settings. The beam properties at the IP, the beam chamber and the detector geometries, the distance between the detector edge and the beam centre were taken into account. This distance was set to 15 $\sigma$ for collision optics and to 10 $\sigma$ for *high-$\beta^*$* ones ($\varepsilon = 3.75$ $\mu$m·rad was used) and a 0.3 mm of dead material was assumed.

For the collision optics the region of high acceptance (> 80%) is limited by $p_T < 0.5$ GeV and $0.06 < \xi < 0.12$, which is significantly smaller than in case of the AFP detectors. The picture changes drastically when *high-$\beta^*$* optics is considered, as these settings are optimised for the elastic scattering measurement in which access to low $p_T$ values for $\xi = 0$ is crucial. One should also note that the limit on the minimum value of the proton $p_T$ decrease with the increase of the $\beta^*$ value. In other words, the higher the $\beta^*$ is, the smaller $t$ values are reachable. It is worth mentioning that the lower value of accessible $p_T$ depends strongly on the distance between the beam centre and the detector edge (see[14] for the reference).

## 6. COLLIMATORS

During the LHC runs with low intensity bunches the two collimators (TCL4 and TCL5), which are installed before the ATLAS forward proton detectors, are kept wide open. Therefore, the geometric acceptance for all measurements, which are planned to be done using the ALFA detectors, is not affected. This situation is different for the high intensity runs that are a part of the AFP physics programme – due to the danger of magnet quenching, the collimators are closed at 15 $\sigma$[¶]. As it is illustrated in Figure 10, such setting would reduce the AFP acceptance by a lot.

To make the measurements possible, the AFP Group proposed[11] to install additional TCL6 collimators in front of the Q6 magnets. This new collimating scheme assumes that the positions of TCL4 and TCL5 will be at the distance of 30 $\sigma$ (15.77 mm) and 50 $\sigma$ (14.53 mm) from the beam, respectively. In addition, the TCL6 new collimator is planned to be positioned at 40 $\sigma$ from the beam. This solution allows to keep a good acceptance for diffracted protons and was admitted as a possible alternative to the present scheme by the LHC Vacuum group.

## 7. SUMMARY

In this paper the LHC apparatus in the vicinity of the ATLAS Interaction Point was described. The proton behaviour in the LHC magnets between the IP1 and ATLAS forward detectors was discussed. The beam size and angular spread at the Interaction Point for various optic modes were calculated. It was shown that the beam size decreases with increasing value of the betatron function (as $\sqrt{\beta^*}$). On the other hand, the beam momentum divergence decreases and is of about hundreds MeV for the *collision* optics and hundreds of keV for the *high-$\beta^*$* ones.

Secondly, it was shown that for the AFP detectors diffractive protons are bent outside the LHC ring (and in $+y$ direction) for all considered optics modes. On the other hand, for the ALFA stations and the *high-$\beta^*$* optics, diffractive protons were going inside the LHC ring.

For the AFP detectors the region of high acceptance (> 80%) is limited by:

---

[¶]The 15 $\sigma$ distance is equal to 7.89 mm in case of TCL4 and 4.36 mm in the case of TCL5.

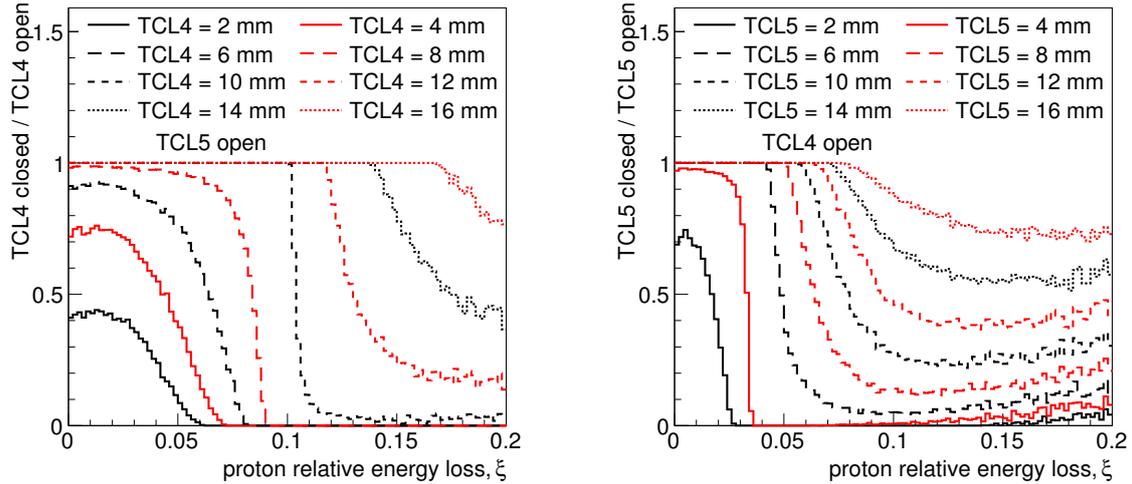

Figure 10. Ratio of the number of events that could be detected with the AFP detector when a given collimator is closed to the number of events when it is wide open as a function of the proton relative energy loss. The default distance of 15 $\sigma$ is equal to 7.89 mm in case of TCL4 and 4.36 mm in the case of TCL5.

- $p_T < 3$ GeV and $0.02 < \xi < 0.12$ for the *collision* optics,
- $p_T < 1$ GeV and $0.07 < \xi < 0.17$ for $\beta^* = 90$ m,
- $p_T < 1$ GeV and $0.1 < \xi < 0.17$ for $\beta^* = 1000$ m.

It is worth to mention that there is almost no acceptance for the optics with $\beta^* = 2625$ m. On the other hand, the specially designed *high-$\beta^*$* optics allows the measurement of protons scattered elastically ($\xi = 0$) with in the ALFA detectors. Moreover, the lower limit of visible transverse momentum decreases with the increase of the value of the betatron function.

Finally, the impact of the LHC collimators on the geometric acceptance was discussed. It was stated that since they are wide open during the low intensity runs, they will not affect the measurements. This situation is was shown to be different during high intensity runs, which are a part of the AFP physics programme. This triggers a need to install new collimator in front of Q6 magnet.

# REFERENCES

[1] ATLAS Collaboration, "*Observation of a new particle in the search for the Standard Model Higgs boson with the ATLAS detector at the LHC.*" Physics Letters B **716** (2012) 1.
[2] CMS Collaboration, "*Observation of a new boson at a mass of 125 GeV with the CMS experiment at the LHC.*" Physics Letters B **716** (2012) 1.
[3] A. Breskin (ed.), R. Voss (ed.), "*The CERN Large Hadron Collider: Accelerator and Experiments.*" J. Instrum. **3** (2008) S08001.
[4] M. Trzebinski, "*Study of QCD and Diffraction with the ATLAS Detector at the LHC.*" CERN-THESIS-2013-166.
[5] K. Steffen, "*High Energy Beam Optics.*" Interscience Monographs and Texts in Physics and Astronomy vol. XVII, 1964.
[6] ATLAS Luminosity and Forward Detector Community, "*ALFA Technical Design Report.*" ATLAS TDR **018**, CERN/LHCC/2008-004.
[7] J. Baechler, O. Kenneth, "*TOTEM-CMS consolidation and upgrade project.*" presented at *LHC Working Group on Forward Physics and Diffraction*, Trento, 2014.
[8] The LHC Study Group collaboration, "*Design study of the Large Hadron Collider (LHC): a multiparticle collider in the LEP tunnel.*" CERN-AC-91-03.

[9] http://proj-lhc-optics-web.web.cern.ch/proj-lhc-optics-web/.
[10] ATLAS Collaboration, "*The ATLAS Experiment at the CERN Large Hadron Collider.*" J. Instrum. **3** (2008) S08003.
[11] L. Adamczyk *et al.*, "*AFP: A Proposal to Install Proton Detectors at 220 m around ATLAS to Complement the ATLAS High Luminosity Physics Program.*" ATLAS-COM-LUM-2011-006.
[12] F. Schmidt, "*MAD-X User's Guide.*" CERN 2005.
[13] E. Forest, F. Schmidt, E. McIntosh, "*Introduction to the Polymorphic Tracking Code.*" CERN.SL.2002.044 (AP) KEK-Report 2002-3.
[14] M. Trzebinski, R. Staszewski, J. Chwastowski, "*LHC High $\beta^*$ Runs: Transport and Unfolding Methods.*" ISRN High Energy Physics, vol. 2012 (2012) 491460.